\documentclass[12pt]{revtex4} 
\usepackage{graphicx}                 
\usepackage{color}                    

\def\bea{\begin{eqnarray}}

\def\eea{\end{eqnarray}}

\def\be{\begin{equation}}

\def\ee{\end{equation}}
\def\nn{\nonumber}

\def\G{\Gamma}

\def\lb{\lambda}

\def\m{\mu}
\def\n{\nu}

\def\r{\rho}
\def\s{\sigma}

\def\t{\tau}

\def\a{\alpha}

\def\b{\beta}

\begin{document}

\title{The Measure in Euclidean Quantum Gravity \footnote{condensed version of a book chapter}}
\author{\sc Arundhati Dasgupta\footnote{arundhati.dasgupta@uleth.ca}}
\affiliation{
Department of Physics and Astronomy,\\University of Lethbridge, Lethbridge, Canada T1K 3M4}

\begin{abstract}In this article a description is given of the measure in Euclidean path-integral in 
quantum gravity, and recent results using the Faddeev-Popov method of gauge fixing. The results 
suggest that the effective action is finite and positive.
\end{abstract}

\maketitle

\section{Introduction}

The gravitational path integral is defined as the sum over all possible
metrics which exist given the boundary data, with each metric history `weighted'
with $e^{\iota/\hbar S}$ ($S$ is Einstein's action). This sum or the integral is the quantum propagator which measures the probability of 
the space-time to evolve from one boundary to the other.  There are several technical difficulties which arise in computing the path-integral 
including the non-linearity of the Einstein action, and the fact that its Euclidean version is unbounded from below. The measure in the integral 
also has to be treated carefully. The Einstein action is diffeomorphism invariant and therefore the measure overcounts the geometries which are 
diffeomorphically related. The straight forward computation of the path-integral involves an evaluation of the physical measure by factoring 
out the diffeomorphism group \cite{hawk}. 
The computation of the scalar sector of the measure is discussed in this article.
This article is a follow up to \cite{adg1} and clarifies some questions raised by a referee and also by R. Loll. In this article we also discuss 
the scalar measure in orthogonal gauge to verify the calculations of \cite{adg1}.
 
In section II a basic description of the path-integral is given and the unboundedness of the Euclidean action is discussed. In section III the measure 
is evaluated and the scalar sector of the measure is exactly computed. It is shown that this calculation takes care of the potential divergence of the 
path-integral due to the unboundedness of the Euclidean action discussed in the previous section.  Section IV is a conclusion.  

\section{The Gravity Path-Integral}

In exact analogy to Feynman's initial formulation, a path-integral for a space-time to evolve from the boundary metric $h^{1}_{\m \nu}$ defined on a 
initial surface
$\Sigma_1$  to another metric $h^2_{\m \nu}$ defined on a final surface $\Sigma_2$ is given by
\be
K(h^1_{\m \nu} \Sigma_1, h^2_{\m \nu} \Sigma_2)= \int {\cal D} g_{\m \nu} \exp\left(\frac{i}{16\pi G\hbar}\int d^4x \sqrt{ g} \ R\right)
\ee
where Einstein's action defined using the determinant $g$ and the scalar Ricci Curvature $R$ of metrics appears in the exponent in the integrand.  However, despite the simplicity of its definition, the actual evaluation of the path-integral has proved very difficult to compute. 

\noindent
(i)Analytic continuation in time does not make sense as `time' coordinate is not uniquely identified in a diffeomorphism invariant
theory.\\
(ii)The gravitational action is non-quadratic and even non-polynomial due to the $\sqrt{ g}$ term, and we can compute
integrals which are essentially Gaussian.\\
(iii)Perturbation about a classical metric with well defined time and metric does not make sense as Feynman diagrams diverge at each order in the perturbation theory and the theory is non-renormalisable.\\
(iv)If one formulates a Euclidean path-integral ab-initio, one finds that the gravitational path-integral is unbounded from below,
and the exponent or the probability for certain configurations diverges making the `integral' including those configurations
meaningless.\\
(v)The gravitational action is diffeomorphism invariant, and hence geometries with the same action are over counted in the measure.\\
Several physicists have been working to solve these issues, or formulate quantum gravity using new approaches. In these article I shall 
 compute the scalar sector of the Faddeev-Popov measure and in the process examine the resolution of the unboundedness of the Euclidean action in the gravitational path-integral.\\ 
 As the `Wick Rotation' from Lorentzian to Euclidean metrics is difficult to implement, one can work with Euclidean or sum over positive definite metrics
or compute the Euclidean path-integral.
\subsection{Faddeev-Popov Measure}
The Euclidean gravitational action as formulated by Einstein $S= -\frac{1}{16 \pi G} \int \sqrt{ g}\ R$ is invariant under transformations
of the metric $g'_{\m \nu} = \frac {\partial x^{\lb}}{\partial x'^{\m}}\frac{\partial x^{\r}}{\partial x'^{\nu}} g_{\lb \r}$ the general coordinate transformations or diffeomorphisms of the manifold. In a infinitesimal transformation, implemented thus $x^{\m}\rightarrow x^{\m} + \xi^{\m}(x^{\m})$, the $\xi^{\m}$ are
the vector fields, the generators of the diffeomorphisms. Thus configurations which are related by diffeomorphisms leave the
action invariant, and represent redundant or unphysical degrees of freedom. The measure however counts them as distinct geometries, and one
has to factor out the diffeomorphism group. The path-integral written in terms of physical degrees of freedom is thus
\be
\int \frac{{\cal D} g_{\m \nu}}{Diff(M)}\exp \left(-\frac{1}{\hbar}S\right)
\ee
where $Diff(M)$ is the diffeomorphism group of the underlying manifold M. Apart from the diffeomorphism group, there also exists 
another set of transformations of the metric, the `conformal transformations'. 
\be
g_{\m \nu} = \exp(2 \phi) \bar g_{\m \nu}
\label{conf}
\ee
these of course leave neither the metric nor the action invariant, (except for special cases where conformal killing orbits leave the metric invariant). 
In \cite{hawk} the Einstein action was shown to be unbounded from below due to the conformal 
mode of the metric. We will discuss this in details in this article.

To isolate the physical measure, it is always useful to talk about 
DeWitt superspace, which is a manifold comprising of points which are metrics. Taking one particular slice of this manifold identifies a unique set of metrics, and rest of the superspace can be obtained by using the diffeomorphisms and/or conformal orbits. 
Thus, in a infinitesimal neighborhood of the slice, one can use the cotangent space element $\delta g_{\m \nu}= h_{\m \nu}$,
and write a coordinate transformation
\be
h'_{\m \nu}= h^{\perp}_{\m \nu} + \nabla_{\m}\xi_{\nu} + \nabla_{\nu}\xi_{\m} + 2\phi g_{\m \nu}
\ee
The $h^{\perp}$ represents the traceless sector of the theory and lies on the gauge slice (the conformal mode has been factored out from this mode of the metric and written isolated as the third term), a diffeomorphism 
generated by $\xi^{\mu}$ and a conformal transformation generated by $\phi$. This is further written in a more useful way
\bea
h'_{\m \nu}& = & h^{\perp}_{\m \nu} + (\nabla_{\m}\xi_{\nu} + \nabla_{\nu}\xi_{\m} -\frac12 \nabla.\xi g_{\m \nu}) + (2\phi + \frac{1}{2} \nabla.\xi)g_{\m \nu} \nn\\
&=&h^{\perp}_{\m \nu} + (L\xi)_{\m \nu} + (2\phi + \frac12 \nabla.\xi)g_{\m \nu}
\eea

where clearly, the trace of the diffeomorphisms is also isolated as it contributes to the conformal sector of the theory and the operator $L\xi$ maps vectors to traceless
two tensors, and adds to $h^{\perp}_{\m \nu}$. The implementation of this coordinate transformation in the measure in the cotangent space of metrics leads to the identification
of a Jacobian, which is the Faddeev-Popov determinant.
\be
{\cal D} h_{\m \nu}= {\cal D} h^{\perp}_{\m \nu}\ {\cal D}\xi_{\mu}\ {\cal D}\phi \ {\rm det} (M)
\ee
The same determinant appears in the base space of the metrics, and using this the 
measure in the path-integral is written in terms of the physical metrics and the Jacobian, and the diffeomorphism group is factored out. 
\be
\int\frac{{\cal D} g_{\m \nu}}{Diff(M)} e^{-S/\hbar}=\int {\cal D} g^{\rm phys}_{\m \nu}\ \frac {{\cal D}\xi_{\mu}}{Diff(M)} {\cal D}\phi {\rm det} (M) e^{- S/\hbar}=\int {\cal D} g^{\rm phys}_{\m \nu}\ {\cal D}\phi \ {\rm det}(M) e^{- S/\hbar}
\ee

The path integral is thus:
\be
\int {\cal D}g_{\m \nu}^{\rm phys} {\cal D}{\phi} (\rm det M) \exp\left( - S/\hbar \right)
\ee
The Faddeev-Popov determinant can be calculated formally as a functional determinant, however, evaluating the determinant as a function of the physical metric
is a task which requires regularisation and appropriate use of new techniques.  A non-perturbative evaluation of the determinant exists for the scalar sector of the determinant \cite{adg1}. We will discuss some subtleties in that calculation in the next few sections.

\subsection{The unboundedness of the Euclidean action}
The Euclidean gravitational action or the action for positive definite metric comprises of 
\be
S= -\frac{1}{16\pi G} \int \sqrt{ g}~ R  ~d^4x 
\ee
If one rewrites the action in terms of the conformal decomposition of the metric (\ref{conf}) 
then the gravity action reduces to
\be
S= -\frac{1}{16 \pi G} \int  d^4 x  \ e^{2\phi} \sqrt{\bar g} \left[\bar R + 6 (\bar\nabla\phi)^2\right].
\ee
The kinetic term of the conformal mode is positive definite, and hence the Euclidean action can assume as
negative values as possible for a rapidly varying conformal factor. This pathology can be assumed to be a signature of presence of redundant degrees of freedom, but in case of Einstein gravity the conformal transformation is not a symmetry of the action and thus the `redundancy' which this represents is not obvious. However, in this paper we investigate the path-integral written in terms of physical variables and examine the case if the negative infinity is indeed cancelled.

Does the conformal mode however uniquely isolate any divergence in the Euclidean action?  The $\sqrt{\bar g}e^{2\phi}\bar R$ term in the action has the `kinetic terms' of the conformally equivalence class of metrics. For compact manifolds we can use the Yamabe Conjecture to reduce the $\bar R$ to a constant, and hence this term cannot make the action unbounded by itself, and thus we ignore the contribution. In case of arbitrary manifolds with boundaries, like the case of flat space time, a slightly more careful treatment is required. We take the form of the Ricci scalar given in terms of the derivatives of the metric
\be
\sqrt{\bar g}\bar R= \sqrt{\bar g} [\bar g^{\mu\sigma}\bar g^{\nu\rho} \partial_{\mu}\partial_{\nu}\bar g_{\sigma \rho}-\bar g^{\mu\nu}\bar g^{\alpha\beta}\partial_{\mu}\partial_{\nu}\bar g_{\alpha\beta}]
\label{curv}
\ee
In the above the second term is potentially problematic, to analyse the terms in a algebraic way, we rewrite the second term in terms of the
derivative of the determinant of the metric $\bar g$. This makes the action take the form:
\bea
\sqrt{\bar g}\bar R& = &
\sqrt{\bar g}\left[\bar g^{\b\n} \bar g^{\a \s} \partial_\b\partial_\a g_{\s \n}\right. \nonumber \\ &&+ \left. \frac1{\bar g^2} \bar g^{\a \b}\partial_\a \bar g \partial_\b \bar g + \bar g^{\a \b}\partial_\a \bar g_{\m \n} \partial_\b \bar g^{\m \n} - \frac{\bar g^{\a \b}}{\bar g} \partial_{\a \b} \bar g\right]  
\eea
Thus a potential divergent contribution to the action might come from the second term of the above for rapidly varying $\bar g$ (the other terms in the above donot contribute with a unique sign, and thus in the action integral negative contributions will 
cancel positive contributions instead of summing up to give unbounded behavior). In case of non-compact geometries we might take the $\bar g_{\m \n}$ to be unimodular and then the divergence of the Euclidean action will be concentrated
in the conformal mode term. In case of arbitrary $\bar g_{\m \n}$, the term has the same sign as the conformal mode term, and thus as discussed in this article, any remnant divergence from the $\bar g _{\m \n}$ can be cancelled by the contribution from the measure. However, a more careful analysis of the divergent terms in the Euclidean action are required. For the purposes of this article we will see that the
solution given cures problems due to any divergent Euclidean action.

\subsection{Gauge Fixing}

The Faddeev-Popov determinant can be obtained using a Gaussian Normalisation condition on the measure \cite{bbm}.

\be
1= \int {\cal D}h_{\m \nu} \exp{(- \int \sqrt{g} ~d^4 x \ h_{\m \nu}~G^{\m \nu \r \t}~h_{\r \t})}
\ee
where $G^{\m \nu \r \t}$ is the DeWitt supermetric obtained in terms of the background metric as
\be
G^{\m \nu \r \t}= \frac12 \left(g^{\m\r}g^{\nu \t}+g^{\m \t}g^{\nu \r} + C g^{\m \nu}g^{\r \t}\right)
\ee

This Gaussian normalisation condition is slightly different from the condition in \cite{bbm}, where they have a half in the exponent, and thus it makes a difference to the normalization constant numerics in the Jacobian.

Interchanging the coordinates of the tangent space leads to the determination of a Jacobian (a function of the metric), 
which is then determined as, 
\be
J^{-1}= \int {\cal D} h^{\perp}_{\m \nu}{\cal D}\phi{\cal D} \hat\xi_{\mu}  \exp \left(- \int~ d^4x ~\sqrt{g} \ h_{\m \nu}~ G^{\m \nu \r \t}~ h_{\r \t}\right) 
\label{fadpop}
\ee

The scalar product in the exponent breaks up into
\bea
 &= & \int d^4 x~\sqrt{g}~ {h^{\perp}_{\m \nu}G^{\m \nu \r \t} h^{\perp}_{\r \t}} + \int d^4 x\sqrt{g} \ \ {L_{\m \nu}G^{\m \nu \r \t} L_{\r \t }}
+ \int d^4x \ \sqrt {g}~ 2~h^{\perp}_{\m \nu}G^{\m \nu \r \t}L_{\r \t}
\nn \\ &+ & 16(1+2 C)\int  d^4 x ~\sqrt{g}~ \Omega^2 \nn
\label{gauss}
\eea

The integration over each of the modes gives a determinant. In case of the tensor determinant, it has to be evaluated
using the projection to the gauge fixed modes.
The gauge fixing is obtained by putting a gauge condition on the $h_{\m \nu}$. This 
in case of the covariant gauge choice can be written as
\be
F\circ h_{\nu}=0
\ee
e.g. in the case of the orthogonal gauge \cite{bbm}, 
\be
\nabla^{\mu}h_{\m \nu}=0 \ \ \ \ \ \ F\equiv \nabla^{\mu}
\ee

Using this, the Faddeev-Popov determinant was obtained in \cite{bbm} from (\ref{fadpop}) as
\be
{\rm det}_S (16(1+2C)){\rm det}_V(F\circ F^{\dag})^{-1/2} {\rm det}_V(F\circ L)
\label{detm}
\ee

If we confine ourselves to the Landau gauge, $F^{\dag}= -2 \nabla^{\mu}$
the vector determinant of (\ref{detm}) is
\be
{\rm det}_V (-2 \nabla\circ L)^{1/2}
\ee
This vector determinant is equivalent to the inverse of 
\be
\int {\cal D}\xi e^{-<\xi F\circ L\xi>}
\ee
where $\xi$ is a vector field. 
In case we write the $\xi^{\mu} =\hat\xi^{\mu} + \nabla^{\mu}\tilde\phi$,
 we can isolate a scalar determinant in the above
\be
\int{\cal D}\tilde\phi\  {\rm det}_S (-\nabla^2)^{1/2} e^{-2\int d^4 x \tilde\phi \ \nabla^{\mu}\nabla^{\nu}L_{\mu\nu\lambda}\nabla^{\lambda}\tilde\phi}
\label{scalar1}
\ee
The scalar sector of the Faddeev-Popov determinant is thus
\be
{\rm det}_S(-\nabla^2)^{-1/2}{\rm det}_S [ 8 (1+2C)\left(-2(\nabla)^4 + 4\nabla_{\mu}\nabla^2\nabla^{\mu} + 4 \nabla_{\mu}\nabla_{\rho}\nabla^{\mu}\nabla^{\rho}\right)]^{1/2} 
\label{scalar}
\ee
 $C$ is the constant in the De-Witt metric
which determines the signature. Where we have absorbed a factor of 
two in the prefactor in the operator expression to retain the conventions of \cite{adg1}. The addition of ${\rm det}_s (-\nabla^2)^{1/2}$ is new in this calculation,
this factor was ignored in the initial version of \cite{adg1} in the transformation to the scalar mode of the diffeomorphism generator. 

\section{Evaluation of the Scalar Measure}
One has to compute the functional determinant (\ref{scalar}) using known techniques like the heat kernel equation. We ignore the finite scalar determinant which multiplies the determinant of the fourth order functional operator in the subsequent discussion. It changes some constants in the effective action. This calculation was done in \cite{adg1} and we clarify the calculation
in this article.

We use the heat kernel regularisation which is defined in order to achieve a $\zeta$ function regularisation of the determinant. Given an operator $F$ with eigenvalues
$\lambda$, the zeta function of p is defined to be
\be
\zeta(p)= \sum_{\lambda}\frac1{\lambda^{p}}
\ee
Clearly, the determinant of the operator, would be given by

\be
\rm det(F)= e^{\rm Tr ln F}
\ee
 
and thus

\be
\rm det(F) = e^{-\zeta'(0)}.
\label{zeta}
\ee
 
Other regularisation schemes can also be used
if required, thus the exact answer would be particular to the way of regularisation. From (\ref{zeta}), the Faddeev-Popov determinant can be written as
an exponential and adds to the classical action in the `weight' of the path-integral creating an effective action. Since the Faddeev-Popov has the 
square root of the determinant appearing in the measure, the exact terms which appear in the effective action for the scalar determinant
is $\G_{\rm trace}= \frac12 \zeta'(0)$. Thus one has to find $\zeta'(0)$ for the scalar determinant of (\ref{scalar}).

The zeta function for a given operator is appropriately written in terms of the `Heat Kernel'.  
\be
\zeta(p)= \frac{\mu^{2 p}}{\Gamma(p)} \int dt\  t^{p-1} {\rm Tr} \exp(-t F)(x,x') 
\ee
The Heat Kernel is precisely the term $U(t, x,x')=\exp(-t F)(x,x')$, $t$ is a parameter and $x,x'$ represent coordinates of the manifold in 
which the operator is defined. The heat kernel satisfies a diffusion differential equation, and can be solved term by term of a series 
$\Sigma a_n t^n$ \cite{dewit}. In \cite{adg1} the operator was slightly transformed by scaling the Heat Kernel by $\exp(-t Q)$, Q being
a function of curvature \cite{adg1} 

The zeta function and its derivative are thus: 
\bea
\zeta (p) &= & \frac{\mu^{2p}}{16 \pi^2 \Gamma(p)} \int dt \ t^{p-3} \ {\rm Tr}_{x} e^{-t Q} \ \sum_{i}  a_i t^i \\
16 \pi^2\zeta'(p) &=& \frac{\mu^{2 p}}{\Gamma(p)} \ \left(\ln \mu^2 - \frac{\Gamma'(p)}{\G(p)}\right) \ \int dt t^{p-3} \ {\rm Tr}_x e^{-t Q} \ \sum_{i}  a_i t^i \nn \\
&+& \frac{\mu^{2p}}{\G(p)} \ \int dt \ t^{p-3} \ \ln t \ {\rm Tr}_x \ e^{-t Q} \ \sum_i a_i t^i 
\label{eq:deriv}
\eea

The finite term as $p\rightarrow 0$ appears in the last term of the derivative of the zeta function (as we know
this might not be the unique the way to extract the finite term,
 this is given in details in \cite{adg1}). The finite term (regulator independent) is remarkably proportional to $a_1$  and is obtained from (\ref{eq:deriv}) in \cite{adg1}
(using a regularisation $\zeta(0)=-1/2$.)
\be
\zeta'(0)_{\rm finite}= -\frac{1}{32 \pi^2} {\rm Tr}_x a_1 
\ee
 
Thus in the effective action $ \G_{\rm trace} + \G_{\rm classical}= \frac12 \zeta'(0) + \G_{\rm classical}$ we get to first order $-\frac{1}{64 \pi^2}{\rm Tr}_x a_1 + \G_{\rm classical}$. In the subsequent
discussion, we fix the $a_1$ in the weak gravity and strong gravity regimes and find the $\G_{\rm trace}+ \G_{\rm classical}$.

These regimes are identified in the following way, the scalar operator in (\ref{scalar}) is taken thus
\be
-2 \nabla^4 + 4\nabla_{\mu}\nabla^2\nabla^{\mu} + 4 \nabla^{\mu}\nabla^{\rho}\nabla_{\mu}\nabla_{\rho}
\ee
By using the commutation relations, one obtains
\bea
&&6\nabla^4 + 4\nabla_{\mu}[\nabla^2,\nabla^{\mu}] + 4 \nabla^{\mu}[\nabla^{\rho},\nabla_{\mu}]\nabla_{\rho}\\
&&6 \nabla^4 + 8 \nabla_{\mu} R^{\mu \nu}\nabla^{\nu} \label{rew}
\eea

We take two limits, one where there is weak field gravity, and one obtains $\nabla_{\m} R^{\mu \nu}\sim 0$, here (\ref{rew})
is approximated by
\be
6\nabla^4\left(1+ \frac{4}{3}(\nabla^4)^{-1}\nabla_{\mu}R^{\mu \nu}\nabla_{\nu}\right).
\ee
And where gravity is strong, one obtains
\be
8\nabla_{\mu}R^{\mu \nu}\nabla_{\nu}\left(1+ \frac{3}{4}(\nabla_{\mu}R^{\mu \nu}\nabla_{\nu})^{-1}\nabla^4\right).
\ee
And thus we approximate the scalar determinant by two determinants

\be
\rm det_S \left(8(1+2C) 6\nabla^4\right) \ \ \ \ \ (\nabla^{\m}R_{\m \nu} \sim 0)
\label{operator1}
\ee
and
\be
\rm det_S\left(8(1+2C) 8 \nabla_{\mu}R^{\mu \nu}\nabla_{\nu}\right) \ \ \ \ \ \ \ (\nabla^{\m}R_{\m \nu}\gg 1)
\label{operator}
\ee

These regimes are identified for the entire metric and not the conformally transformed metric. The first one (\ref{operator1}) can be evaluated using the Heat Kernel for the Laplacian for arbitrary space times, under certain
boundary conditions, by splitting the fourth order operator into product of two Laplacians whose heat kernel expansions
are well known \cite{dewit}.
 
(i) $\nabla_{\mu}R_{\m \nu}\approx 0$, the Faddeev-Popov determinant is the square root of 
\be
{\rm det}(8(1+2C)6\nabla^4)= {\rm det}(-8(1+2C)6\nabla^2) {\rm det} (-\nabla^2)
\ee

In the factorised form, the first scalar determinant is a divergent one for $C<-1/2$, and the second one is a convergent one. We discuss the divergent determinant's
contribution to the effective action as this should cancel the divergence from the classical action.

The first scalar determinant is of a Laplacian and thus we use the heat kernel of a Laplacian, and analytically continue to the
divergent regime of $C<-1/2$.
The $a_1$
coefficient of the Laplacian is well known, and to quote \cite{dewit,bar}
\be
a_1= \frac{1}{6} R
\ee
What is interesting is that this term is dimensional and it is obvious that the space of measures
does not have any length scales like the Planck length to make the term dimensionless in the
exponential. To retain the correct dimensions, one had to introduce in the definition of the measure and the determinants a scale.
I scale the coefficient $a_1$ by a appropriate length squared :$ l_{\rm un}^2$ (in \cite{adg1} this had been taken to be Planck length squared) to get the exponential dimensionless and restore the $16\pi G$ in the classical action. Writing $R$ in terms of $\bar R$ and $(\nabla \phi)^2$, and using the constants in the determinant one gets
as the coefficient of the kinetic term of the conformal mode in the effective action ($\G_{\rm trace} + \G_{\rm classical}$)
\be
-\frac{1}{16 \pi}\left[1 + \frac{2(1+ 2C)}{\pi}\frac{l_p^2}{l_{\rm un}^2}\right]
\ee

Thus the positive action takes over at $(1+2C)> - l_{\rm un}^2 \pi/2 l_p^2$. From Einstein' action $C=-2$, and Euclidean Einstein gravity has a positive definite effective action for $l_{\rm un}=l_{p}$. The number $-\pi/2$ might differ for different
regularisation schemes, (and conventions for defining determinants from Gaussian integrals) but it is indeed a finite number, and we should be in the realm of a
convergent path-integral for Euclidean quantum gravity. Note that the effect of this calculation has been in the end change of the
overall sign of the action by a minus sign. This achieves the required `convergent behavior' 
of the exponent in the path-integrand. This contribution from the measure has the right sign, irrespective of what the sign of the classical bare action is. We saw that the classical bare action can be unbounded from below, and this change in sign makes those divergent actions contribute
with convergent weights to the gravitational path-integral. From a premature analysis of the classical action it seems that the wildly divergent terms appear with a negative sign and thus reversing this sign makes the path-integral potentially convergent.

(ii) In the case of the regime $\nabla^{\mu}R_{\m \nu} \gg 1$, the relevant operator is (\ref{operator}) and the coefficient $a_1$ is determined for $-8\nabla^{\mu}R_{\m\n}\nabla^{\n}$ (with a scaling and in anticipation that the (1+2C) prefactor will be negative) from \cite{adg1}

\bea
a_1& =& -2 R_{\lb \s}[(R^{-1})^{\m \nu} \nabla^{\lb} R_{\m \nu}] [(R^{-1})^{\m'\nu'}\nabla^{\s} R_{\m' \nu'}] \nn \\ 
&-& 4 R_{\lb \s} (R^{-1})^{\m \nu}\nabla^{\s}\nabla^{\lb}R_{\m \nu} - 8 R^2
\label{a12}
\eea
We find that 
  $-R^2$ term has exactly the
sign required to cancel the contribution from the $(\bar \nabla\phi)^2$ term in the classical action, as writing R in terms of $\bar R$ and $\bar \nabla \phi$, one finds $(\bar\nabla\phi)^4$ from $R^2$,
which dominate for configurations with rapidly varying $\phi$. The other terms do not give divergent negative terms. Any further negative divergence from $\bar g$ is also rendered positive by this squared term. Thus $\G_{\rm trace} + \G_{\rm classical}$ in this regime emerges as positive definite. 
Note in this non-perturbative regime, the effective action at this order does not merely change by an overall minus sign but has additional non-trivial contributions
proportional to $R^2$ and higher derivative terms, which dominate over the classical bare action.

\section{Conclusion}
In this article we clarified and verified some calculations of \cite{adg1} regarding the scalar measure in the gravitational path-integral. We showed by 
an exact computation in the orthogonal gauge, the scalar operator which 
appears in the measure is the same as obtained in \cite{adg1}. 
This is not surprising as the scalar operator was obtained for the generic case in \cite{adg1} and
should be the same in any covariant calculation. We clarified the fact that the Euclidean action
can be unbounded only from below by isolating the potentially diverging terms in the calculation
and the resolution given in \cite{adg1} though a more careful analysis has to be done. We then explained 
the results of \cite{adg1} in evaluating the scalar measure and used that to observe that the effective action 
in the gravitational action is positive. What is very interesting is the
emergence of finite terms by regularising the determinant in a non-perturbative way. Work is in progress in 
obtaining a regularisation of the entire Faddeev-Popov determinant.


\begin{thebibliography}{99}

\bibitem{hawk} G. W. Gibbons, S. W. Hawking, M. J. Perry, Nucl. Phys. {\bf B 138} 141 (1978).
 J. Hartle, K. Schleich, Phys. Rev. {\bf D 36} 2342 (1987).
K. Schleich, Phys. Rev. {\bf D 39} 2192 (1989).
 P. Mazur, E. Mottola, Phys. Rev {\bf D 64} 104022, 2001, Nucl. Phys. {\bf B 341}:187, (1990). E. Mottola, J. Math. Phys. {\bf 36} 2342 (1987).
 A. Dasgupta, R. Loll, Nucl. Phys. {\bf B606} 357, (2001).
\bibitem{bbm} Z. Bern, M. Blau, E. Mottola, Phys. Rev. {\bf D43}: 1212 , (1991).
\bibitem{adg1} A. Dasgupta, {\it The Gravitational Path-Integral and the Trace of the Diffeomorphisms} Arxiv: 0801.4770 [gr-qc] (to appear in General. Relativ. and Grav.)
\bibitem{dewit}J. B. S. Dewitt, {\it Dynamical theory of groups and fields}, Gordon and Breach Science Publishers, (1965).
I. G. Avramidi, {\it Heat Kernel and Quantum Gravity}, Lecture Notes in Physics, Springer, (2000). 
S. A. Fulling, {\it Aspects of Quantum Field Theory in Curved Space-Time}, Cambridge University Press, (1989).
\bibitem{bar} A. O. Barvinsky, G. A. Vilkovisky, Phys. Rept. {\bf 119} 1, (1985).
\end{thebibliography}
\end{document}